\documentclass[aps,twocolumn,showpacs,preprintnumbers]{revtex4}

\input{epsf}

\def\AJ{{\it Ap. J.} }

\def\CQG{{\it Class. Quantum Gravity} }

\def\GRG{{\it Gen. Relativity and Gravitation} }

\def\IJMP{{\it Int. J. Mod. Phys.} }

\def\NAT{{\it Nature} }

\def\PL{{\it Phys. Lett.} }
\def\PR{{\it Phys. Rev.} }
\def\PRL{{\it Phys. Rev. Lett.} }

\def\al{\alpha} \def\be{\beta}  \def\de{\delta}
   
   \def\ka{\kappa}
\def\la{\lambda}   
\def\si{\sigma}   
   
  \def\De{\Delta} 
\def\La{\Lambda}   
 \def\Om{\Omega} \def\mn{{\mu\nu}}

 \def\frac#1#2{{\textstyle{{#1}\over
{#2}}}} 
\def\lsim{\mathrel{\rlap{\lower4pt\hbox{\hskip1pt$\sim$}}
\raise1pt\hbox{$<$}}}
\def\gsim{\mathrel{\rlap{\lower4pt\hbox{\hskip1pt$\sim$}}
\raise1pt\hbox{$>$}}} \def\sqr#1#2{{\vcenter{\vbox{\hrule height.#2pt
\hbox{\vrule width.#2pt height#1pt \kern#1pt \vrule width.#2pt} \hrule
height.#2pt}}}}
\def\square{\mathchoice\sqr66\sqr66\sqr{2.1}3\sqr{1.5}3}
\def\beq{\begin{equation}} \def\eeq{\end{equation}}
\def\beqa{\begin{eqnarray}} \def\eeqa{\end{eqnarray}}
\def\eq#1{Eq. (\ref{#1})}

\begin{document}

\title{Accelerated expansion from a non-minimal gravitational coupling to matter}

\vskip 0.2cm

\author{O. Bertolami}
\email{orfeu@cosmos.ist.utl.pt}

\author{P. Fraz\~ao}
\email{pedro.frazao@ist.utl.pt}

\author{J. P\'aramos}
\email{paramos@ist.edu}

\vskip 0.2cm

\affiliation{Instituto Superior T\'ecnico, Departamento de
F\'{\i}sica, \\Av. Rovisco Pais 1, 1049-001 Lisboa, Portugal }

\vskip 0.5cm

\affiliation{Instituto de Plasmas e Fus\~ao Nuclear, Instituto Superior T\'ecnico, \\Av. Rovisco Pais 1, 1049-001 Lisboa, Portugal}

\vskip 0.2cm

\vskip 0.5cm

\date{\today}

\begin{abstract}

It is shown that a non-minimal coupling between the scalar curvature and the matter Lagrangian density may account for the accelerated expansion of the Universe and provide, through mimicking, for a viable unification of dark energy and dark matter. An analytical exploration is first performed, and a numerical study is then used to validate the obtained results. 
The encountered scenario allows for a better grasp of the proposed mechanism, and sets up the discussion for improvements that can lead to a 
complete agreement with the observational data.

\vskip 0.5cm

\end{abstract}

\pacs{04.20.Fy, 04.80.Cc, 97.10.Cv \hspace{2cm}Preprint DF/IST-1.2010}

\maketitle

\section{Introduction}\label{intro}

Many modifications of theories of gravity are motivated by one of the outstanding puzzles of modern cosmology: the origin of the observed accelerated expansion of the Universe. The most common approach to this issue relies on the presence of a dominating dark energy component (with $\Om_{DE} \approx 70 \%$) \cite{accexp}, which might arise from several competing candidates: a cosmological constant term in the Einstein-Hilbert action, a scalar field, usually referred to as quintessence \cite{quintessence}, chameleon fields \cite{chameleon}, or other alternatives to General Relativity (GR), such as the cardassian model with a modification of the Friedmann equation \cite{cardassian}, braneworld scenarios \cite{branes} or the generalized Chaplygin gas unification of dark energy and dark matter \cite{Chaplygin}.

Aiming at a description of this dark energy component, several authors have put forward proposals based upon the so-called $f(R)$ models, where a modified action functional exhibiting a non-linear function of the scalar curvature $R$ is considered \cite{fR}. This is usually regarded as stemming from a low-energy phenomenological approximation to some higher energy fundamental theory; indeed, one-loop renormalization of GR requires the introduction of higher order terms in the curvature in the Einstein-Hilbert action functional, and other available invariants --- such as contractions of the Ricci or of the Riemann tensor --- may also arise when quantum corrections arising from string theory are considered (see Ref. \cite{Sotiriou} for a thorough discussion). In a cosmological context, these models usually rely on a decreasing $f(R)$ function that, since the scalar curvature is decreasing, deviates strongly from GR at late times --- thus producing the required accelerated expansion \cite{capoexp}.

Aiming to further extend the $f(R)$ theories, a model was advanced exhibiting not only a non-linear $f(R)$ term in the action functional, but also a non-minimal coupling between the matter Lagrangian density $\mathcal{L}_m$ and the scalar curvature \cite{f2R}. The purpose of this work is to show that this latter model may be used to account for the accelerated expansion of the Universe without any explicit additional matter component ({\it e.g.} scalar fields). By resorting to a previous study where it was shown that this non-minimal gravitational coupling with matter can mimic known dark matter profiles (thus producing the reported flattening galaxy rotation curves) \cite{mimic}, one concludes that the proposed model yields, through its gravitational impact, a viable unified scheme to mimic the 
presence of both dark energy and dark matter.

This work is organized as follows: the non-minimal gravitational coupling model is discussed in Section II; an analytical work, establishing quantitative results related to the accelerated expansion of the Universe, is discussed in Section III; this discussion sets up the numerical calculation that confirms the proposed scenario in Section IV. Finally, the conclusions are presented in Section V.

\section{The model}

Following the discussion of the previous section, one postulates the following action for the theory \cite{f2R}:

\beq S = \int \left[ {1 \over 2}f_1(R) + [1 + \la f_2(R) ] \mathcal{L}_m \right] \sqrt{-g} d^4 x \label{action} ~~,\eeq

\noindent where $f_i(R)$ (with $i=1,2$) are arbitrary functions of the scalar curvature $R$, $\mathcal{L}_m$ is the Lagrangian density of matter and $g$ is the metric determinant. The contribution of the non-minimal coupling of $f_2$ is gauged through the coupling constant $\la$, which has dimensions $[\la] = [f_2]^{-1}$. The standard Einstein-Hilbert action is recovered by taking $f_2=0$ and $f_1= 2 \ka (R - 2 \La)$, where $\ka = c^4 /16 \pi G$ and $\La$ is the cosmological constant.

Variation with respect to the metric $g_\mn$ yields the field equations, here arranged as

\beqa \label{EE0} && \left( F_1 + 2 \la F_2 \mathcal{L}_m \right) R_\mn - {1 \over 2} f_1 g_\mn =  \\ \nonumber &&\De_\mn \left(F_1 + 2 \la F_2 \mathcal{L}_m \right) + \left( 1 + \la f_2 \right) T_\mn~~, \eeqa

\noindent where one defines $\De_\mn = \nabla_\mu \nabla_\nu - g_\mn \square $ for convenience, and writes $F_i (R) \equiv f_i'(R) $, omitting the argument. The matter energy-momentum tensor is, as usually, defined by

\beq T_\mn = -{2 \over \sqrt{-g}} {\de \left(\sqrt{-g} \mathcal{L}_m \right) \over \de g^\mn } ~~. \eeq

\noindent By taking the trace of \eq{EE0}, one obtains

\beqa \label{trace0} && \left( F_1 + 2 \la F_2 \mathcal{L}_m \right) R - 2 f_1 = \\ \nonumber && -3 \square \left(F_1 + 2 \la F_2 \mathcal{L}_m \right) + \left( 1 + \la f_2 \right) T  ~~.\eeqa

The Bianchi identities, $\nabla^\mu G_\mn = 0$ imply the non-(covariant) conservation law

\beq \nabla^\mu T_\mn = {\la F_2 \over 1+ \la f_2} \left( g_\mn \mathcal{L}_m - T_\mn \right) \nabla^\mu R ~~, \label{non-cons} \eeq

\noindent which, in the context of an analogy between \eq{action} and a scalar-tensor theory, may be interpreted as due to an energy exchange between matter and the scalar fields associated with the model \cite{accexp} for the non-trivial $f_1(R)$ and $f_2(R)$ terms \cite{scalar}.

Since a complete study of the joint effect of a non-trivial $f_1(R) $ and $f_2(R) $ is too involved, one focus the attention on the latter, thus setting $f_1(R) = 2 \ka R $ (discarding the cosmological constant $\Lambda$); this reduces \eq{EE0} to

\beqa \label{EE} && \left( 1+ {\la \over \ka} F_2 \mathcal{L}_m \right) R_\mn - {1 \over 2} R g_\mn =  \\ \nonumber && {\la \over \ka} \left( \square_\mn - g_\mn \square \right) \left( F_2 \mathcal{L}_m \right) + {1 \over 2 \ka}  \left( 1 + \la f_2 \right)  T_\mn~~, \eeqa

\noindent and, taking the trace, the equivalent of \eq{trace0},

\beqa \label{trace} && \left( 1 - {\la \over \ka} F_2 \mathcal{L}_m \right) R = \\ \nonumber && 3{\la \over \ka} \square \left( F_2 \mathcal{L}_m \right) - { 1 \over 2 \ka}  \left( 1 + \la f_2 \right) T  ~~.\eeqa

\section{Accelerated expansion phase}

\subsection{Power-law expansion}

One begins by rewriting \eq{EE0} in a more natural way,

\beq G_\mn = {1 \over 2 \ka } \left( T^m_{\mn} + T^c_{\mn} \right) ~~,\label{EEusual} \eeq

\noindent so that, using $L_m = -\rho$ (see \cite{Ld} for a discussion), one defines

\beqa T^m_{\mn} &=& {2 \ka \over F_1 - 2 F_2 \rho} T_\mn ~~, \\ \nonumber T^c_{\mn} &=& {2 \ka \over F_1 - 2 F_2 \rho } \times  \bigg[ \De_\mn (F_1 - 2 F_2 \rho) + \\ \nonumber &&  {1 \over 2} (f_1 - F_1 R) g_\mn + F_2 \rho R g_\mn + f_2 T_\mn \bigg] ~~.\eeqa

One assumes that the matter content of the Universe is described by a perfect fluid, endowed with an energy-momentum tensor

\beq T_\mn = (\rho + p ) u_\mu u_\nu + p g_\mn ~~.\eeq

\noindent Resorting to the Friedmann-Robertson-Walker metric given by the line element below,

\beq ds^2 = - dt^2 + a^2(t) \left( {dr^2 \over \sqrt{1-kr^2}} + d\Om^2 \right) ~~. \eeq

\noindent and considering the isotropy and homogeneity implied by the Cosmological Principle, one writes the four-velocity as $u_\mu =(u_0 , 0,0,0)$. The normalization condition $ u_\mu u^\mu = -1 $ thus translates into $ u_0^2 = 1$, and the energy-momentum components reads

\beqa T_{00} &=& \rho ~~, \\ \nonumber T_{rr} &=& p g_{rr} = a^2 p ~~. \eeqa

It is easy to check that, likewise $T_\mn$ with the adopted four-velocity, $T^c_{\mn}$ is also diagonal: one may write the corresponding curvature ``density'' and ``pressure'',

\beqa \label{rhoc} \rho_c &= & T^c_{00}= \\ \nonumber &&{2 \ka \over F_1 -2 F_2 \rho}  \bigg[ (f_2 - F_2 R )\rho - {1 \over 2 } (f_1 - F_1 R) - \\ \nonumber && 3H \left[ (F'_1 - 2 F'_2 \rho ) \dot{R} - 2 F_2 \dot{\rho} \right] \bigg] ~~,  \eeqa 

\beqa \label{pc} p_c &= & {T^c_{rr} \over a^2}= \\ \nonumber && {2 \ka \over F_1 -2 F_2 \rho} \bigg [ (F'_1 - 2 F'_2 \rho) (\ddot{R} + 2 H \dot{R}) + \\ \nonumber && (F''_1 -2 F''_2 \rho) \dot{R}^2  +F_2 \left[ \rho R -2(\ddot{\rho}+2H \dot{\rho}) \right] +  \\ \nonumber &&  {1 \over 2} (f_1 - F_1 R) -4 F'_2 \dot{\rho} \dot{R} +  f_2 p \bigg] ~~. \eeqa 

Defining the Hubble parameter $H = \dot{a}/a$, leads to the Friedmann equation

\beq \label{friedmann} H^2 + {k \over a^2} = {1 \over 6 \ka } (\rho_m + \rho_c)~~, \eeq

\noindent and the Raychaudhuri equation

\beq \label{raychaudhuri} {\ddot{a} \over a} = \dot{H} + H^2 = -{1 \over 12 \ka } \left[ \rho_m + \rho_c + 3(p_m + p_c ) \right] ~~,\eeq

\noindent the latter providing the condition for an accelerated expansion of the Universe, $ \rho_m + \rho_c + 3(p_m + p_c ) < 0$ .

Defining the deceleration parameter as 

\beq \label{defq} q = -{\ddot{a} a \over \dot{a}^2}~~, \eeq

\noindent one may combine the two preceding equations to obtain 

\beq \label{qeq} q = {1 \over 2} + {1 \over 4\ka} {p_c \over H^2}~~. \eeq

To provide an insight on the effect of the coupling with matter, one may write the curvature pressure and density $p_c$ and $\rho_c$ arising from two separate cases. On one hand, in the absence of matter, $f_2(R) =0$, one has

\beqa \label{c1} \rho_{c} &= & -{2 \ka \over F_1 }  \bigg[  {1 \over 2 } (f_1 - F_1 R) + 3H F'_1  \dot{R} \bigg]  ~~, \\ \nonumber 
p_{c} &=&  {2 \ka \over F_1} \bigg [ F'_1 (\ddot{R} + 2 H \dot{R}) + F''_1 \dot{R}^2  + {1 \over 2} (f_1 - F_1 R)  \bigg]  ~~.\eeqa 

\noindent On the other hand, including matter, $f_2(R) \neq0$, and setting $f_1(R) = 2 \ka R $, leads to

\beqa \label{c2} \rho_{c} &=& {\ka \over \ka - F_2 \rho}  \bigg[ (f_2 - F_2 R )\rho + 6H \left( F'_2 \rho \dot{R} + F_2 \dot{\rho} \right) \bigg] ~~, \\ \nonumber 
p_{c} &=&  {\ka \over \ka - F_2 \rho} \bigg [ - 2 F'_2 \rho (\ddot{R} + 2 H \dot{R}) - 2 F''_2 \rho \dot{R}^2  + \\ \nonumber && F_2 \left[ \rho R -2(\ddot{\rho}+2H \dot{\rho}) \right] -4 F'_2 \dot{\rho} \dot{R} +  f_2 p \bigg]~~. \eeqa 

\noindent There is a clear difference between the expressions above: the latter two explicitly depend upon the matter density and pressure $\rho$ and $p$, while the former two are a function of $R$ (and its derivatives) only. This, of course, stems from the non-minimal coupling between matter and geometry, but implies that one cannot simply neglect the contribution from matter when solving the Friedmann and Raychaudhuri equations. Indeed, in Ref. \cite{capoexp}, this allows for the determination of a relation between the evolution of the scale factor $a(t)$ and the exponent $m$ present in the non-trivial term $f_1(R) = R_1 (R/R_1)^m$; in the present case, although one can still assume that $\rho_m < \rho_c$, $p < |p_c|$, the density and pressure appear in the above definitions, so that this does not translate directly into setting $\rho= p = 0$.

In order to solve \eq{friedmann} and (\ref{raychaudhuri}), one assumes a flat $k=0$ scenario and inserts the {\it Ansatz} $a(t) = a_0 (t/t_0)^\be$ for the evolution of the scale factor --- physically interesting since it gives rise to a constant deceleration parameter, with $\be>0$ for an expanding Universe and $\be>1$ for accelerated expansion. Thus,

\beqa H &\equiv&  {\dot{a} \over a} = {\be \over t} ~~, \\ \nonumber \label{scalarcurvature} R & \equiv&  6 \left[ \left({\dot{a} \over a}\right)^2 + {\ddot{a} \over a} \right] = 6(\dot{H} + 2 H^2) = {6\be \over t^2} (2\be -1)~~, \\ \nonumber q & \equiv& -{\ddot{a}a \over \dot{a}^2} = {1 \over \be } -1~~. \eeqa

Since one is interested in studying the effect of the non-minimal coupling $f_2(R)$, one assumes that it dominates any non-trivial addition to the usual linear curvature term, thus one can set for simplicity $f_1(R)= 2\ka R$. Given the assumed power-law expansion, the former is assumed to have the form $f_2(R) = (R/R_2)^n$, prompting for the search of the relation between the exponents $n$ and $\be$, as well as the physical meaning of the coupling strength $R_2$.

Furthermore, one requires some foreknowledge of the evolution of the matter density  $\rho$ and pressure $p$; one may assume that it is modelled as a dust distribution with $p= 0$, and looks at \eq{non-cons} for the evolution of $\rho$. Luckily, although the non-(covariant) conservation of the energy-momentum tensor is perhaps the most striking fundamental implication of the model here studied, the introduction of  the adopted energy-momentum for a perfect fluid and the Lagrangian density $\mathcal{L}_m= -\rho$ yields a vanishing {\it r.h.s.} for \eq{non-cons}. 

Following the standard interpretation, one can state that the expansion of the Universe remains adiabatic, with no direct transfer of energy between matter and ``curvature'' component expressed in \eq{c2}. The $\nu = 0$ component of \eq{non-cons} reads:

\beq \dot{\rho} + 3H \rho = 0 \rightarrow \rho(t) = \rho_0 \left({a_0 \over a(t)}\right)^3 = \rho_0 \left({t_0 \over t} \right)^{3\be} ~~. \eeq

With the above law for $\rho(t)$ and the expressions for $f_2(R)$ and $a(t)$, the curvature density $\rho_c$ and pressure $p_c$ read

\beqa \rho_c & =& 6 \ka \rho_0 \be \times \\ \nonumber && \left[{  1-2\be + n(5 \be + 2n -3)  \over n  \left({t\over t_0}\right)^2 \rho_0 - \left({t\over t_0}\right)^{3\be} \left({t \over t_2}\right)^{2n} \left[6\be (2\be-1)\right]^{1-n} \ka}\right] ~~, \eeqa

\noindent and

\beqa p_c &=& 2 \ka \rho_0 n \times \\ \nonumber &&   \left[{ 2+4n^2 -\be (2+3\be) +n(8\be -6) \over n  \left({t\over t_0}\right)^2 \rho_0 - \left({t\over t_0}\right)^{3\be} \left({t \over t_2}\right)^{2n} \left[6\be (2\be-1)\right]^{1-n} \ka }\right] ~~. \eeqa

\noindent defining $t_2 \equiv R_2^{-1/2}$, for simplicity.

In what follows, one assumes alternatively that the denominator of the curvature pressure and density is dominated by either $F_2 \rho$ or the constant $\ka$; for simplicity, the regime $F_2 \rho > \ka$ is dubbed ``$+$ regime'', with the converse leading to the ``$-$ regime''. Actually, one sees that

\beq F_2 \rho = n \left({R \over R_2}\right)^n {\rho \over R} = n \left[ 6\be(2\be-1)\right]^{n-1} \rho_0 {t_2^{2n} t_0^{3\be}  \over t^{2(n-1)+3\be}}  ~~, \label{F2rho} \eeq

\noindent thus scaling as $t^{2-2n-3\be}$.

\subsubsection{The $+$ regime: $F_2 \rho > \ka$}

Clearly, both the curvature pressure $p_c$ as well as the density $\rho_c$ experience two separate time evolutions, signaled by the relevance of $F_2(R) \rho $ on the denominator of \eq{c2}.

One first attempts to solve the Friedmann and Raychaudhuri Eqs. (\ref{friedmann}) and (\ref{raychaudhuri}) in the regime $F_2 \rho > \ka$, so that the curvature pressure and density are given approximately by

\beqa \rho_c & =& {6 \ka \be \over t^2} \left( {1 - 2\be \over n} + 5\be + 2n -3 \right) ~~, \\ \nonumber p_c & = & {2\ka \over t^2} \left[2+4n^2 -\be (2+3\be) +n(8\be -6)  \right]  ~~. \eeqa

\noindent Inserting this into the Friedmann \eq{friedmann} leads to $\be=\be_+ \equiv (1-n)/2$, which trivially satisfies the Raychaudhuri \eq{raychaudhuri}. The condition for an expanding Universe $\be_+>0$ yields the constraint $n<1$.

Replacing onto \eq{F2rho} leads to $F_2 \rho \propto t^{\be_+}$; since $\be_+ >0$, one concludes that, once the inequality  $F_2 \rho > \ka$ sets in, the {\it l.h.s.} increases with time: the $+$ regime, once attained, remains valid.

\subsubsection{The $-$ regime: $F_2 \rho < \ka$}

The analysis of regime $F_2 \rho < \ka$ is slightly lengthier, since the value of $\rho_0$ appears explicitly in the approximated expressions for curvature density and pressure, as can be seen below:

\beqa \rho_c & =& -6 \rho_0 \be {  1-2\be + n(5 \be + 2n -3)  \over  \left({t\over t_0}\right)^{3\be} \left({t \over t_2}\right)^{2n} \left[6\be (2\be-1)\right]^{1-n} } ~~, \\ \nonumber p_c &=& -2 \rho_0 n { 2+4n^2 -\be (2+3\be) +n(8\be -6) \over  \left({t\over t_0}\right)^{3\be} \left({t \over t_2}\right)^{2n} \left[6\be (2\be-1)\right]^{1-n} }~~.   \eeqa

\noindent Since the {\it l.h.s.} of the Friedmann \eq{friedmann} falls as $t^{-2}$, the exponent $\be$ can be directly obtained from the expression for the curvature density,

\beq 3\be + 2n = 2 \rightarrow \be = \be_-(n) \equiv {2 \over 3}(1-n)~~. \eeq

\noindent Inserting this back into the Friedmann equation yields the value for the initial density,

\beq \label{rho-} \rho_0 ={8 \over 3} \left({3 \over 4}\right)^n (1-n) (1-5n+4n^2)^{-n} \left({t_0 \over t_2}\right)^{2n} {\ka \over t_0^2} ~~. \eeq

\noindent As before, the above expressions trivially satisfy the Raychaudhuri \eq{raychaudhuri}. The condition for an expanding Universe $\be_- > 0$ also leads to the upper bound $n<1$.

Since $3\be_- = 2(1-n)$, \eq{F2rho} indicates that $F_2 \rho $ is constant. Therefore, one concludes that if the $-$ regime is attained it is also permanent.

\subsubsection{Regime validity}

In the previous paragraphs, one has concluded that there are two possible regimes where the effect of the non-minimal coupling $f_2$ is dominant, corresponding to the positive or negative sign of $F_2 \rho -\ka$. Furthermore, it was shown that the $+$ regime leads to an increasing $F_2 \rho$ term, so that the corresponding inequality $F_2 \rho > \ka$ becomes even stronger; likewise, the $-$ regime yields a constant $F_2 \rho $ term, so that $F_2 \rho $ remains smaller than $\ka$.

If there are no possible transitions between the two regimes, there is still the issue of the onset of the dominance of the non-minimal coupling, that is, the condition $f_2(R) > 1$. Since one adopts the power-law form $f_2(R)=(R/R_2)^n $ and the scalar curvature $R$ decreases and one aims for a late-time dominance leading to the currently observed accelerated expansion, one concludes that the exponent $n$ must be negative.

This requirement for a negative exponent is, in essence, analog to the one found in a previous study concerning a mimicking mechanism for dark matter \cite{mimic}: in that work, an inverse power-law was required so that the effects of the non-minimal coupling become dominant at large distances, thus leading to the flattening of the rotation curves of galaxies. Conversely, the application of the considered model to astrophysical objects with high densities (such as the Sun \cite{sun}) leads one to the consideration of a linear coupling $f_2(R) \propto R$.

Since $R = 6\be(2\be-1)/t^2$, it is clear that the non-minimal coupling dominates after a transition time $T = \sqrt{6\be(2\be-1)} t_2$ (when $R<R_2$); considering that $\be$ is of order unity ($\be = 2/3$ for the matter dominated phase), this translates into $T \sim t_2$. Thus, the issue of evaluating the sign of $F_2 \rho - \ka$ simplifies to the direct evaluation of this quantity at $T=t_2$. To do this, one requires the parameters $t_0$ and $\rho_0$ specifying the evolution of the matter density: it suffices to consider the WMAP7 value $t_0 = 13.73 ~Gy$ and $\rho_0 = \Om_m \rho_{crit}$, with $\Om_m \sim 0.3$ the relative matter density in the Universe and $\rho_{crit} = 3H_0^2 /8 \pi G \sim 10^{-26}~kg/m^3 $ its critical density, from Ref. \cite{WMAP7}; replacing $\ka = 1/(16\pi G)$ yields $\rho_0 t_0^2 /\ka \sim 0.1$, thus

\beqa \label{which} {1 \over \ka} \left| F_2 \rho  \right|_{t=t_2} &=& n \left[ 6\be(2\be-1)\right]^{n-1} {\rho_0 t_0^2 \over \ka}\left({t_0 \over t_2 }\right)^{3\be -2} \simeq \\ \nonumber && 0.1 n \al^{n-1} \left(4n^2 -5n +1\right)^{n-1} \left({t_0 \over t_2 }\right)^{3\be -2} ~~. \eeqa

\noindent where $\al$ signals the alternative regimes, $\al = 1$ for the $+$ regime or $\al =  4/3$ for the $-$ regime.

Since $\be > 2$ (marking an accelerated expansion of the Universe), one concludes that an earlier onset of the non-minimal coupling dominance (that is, a smaller value of $t_2$) increases the {\it r.h.s.} of the above expression --- eventually leading to the condition $F_2 \rho / \ka > 1$ that marks the $+$ regime. Indeed, there is an interplay between the two terms of the above expression: the term $0.1n[4(4n^2-5n+1)/3]^{n-1} $ is smaller (in absolute value) than approximately $5.6 \times 10^{-3}$ in the domain $n<1$; however, since the accelerated expansion has already begun, $t_2 < t_0$ and $\be > 1$, thus leading to $(t_0/t_2)^{3\be-2} > 1$.

The issue is settled by resorting to observational data in order to fix the accelerated expansion onset time $t_E \sim t_2$. This may be written in terms of the redshift $z_E$ marking the change of sign of the deceleration parameter $q$, through

\beq  \left({t_0 \over t_E}\right)^{3-2\be} \sim  \left({t_0 \over t_2} \right)^{3\be -2}= \left({a_0 \over a(t_2)}\right)^{3-2/\be} = \left(1+z_E\right)^{3-2/\be} ~~. \eeq

Taking the best-fit value $z_E \approx 0.36$ \cite{qz0} and inserting into \eq{which}, together with the expressions for $\al$ and $\be_-(n)$, $\be_+(n)$, yields

\beqa  \label{cases}
&&{1 \over \ka} F_2 \rho = 0.1 n \left(4n^2 -5n +1\right)^{n-1} \times \\ \nonumber
&& \cases{ \left(1.36\right)^{(1+3n)/(n-1)}  & {\rm , }$ F_2 \rho > \ka$
\cr \left(1.36\right)^{3n/(n-1)} (4 / 3)^{n-1}   & {\rm , }$ F_2 \rho < \ka$}~~.
\eeqa

\begin{figure}[tbp]
\centering
\epsfxsize=\columnwidth
\epsffile{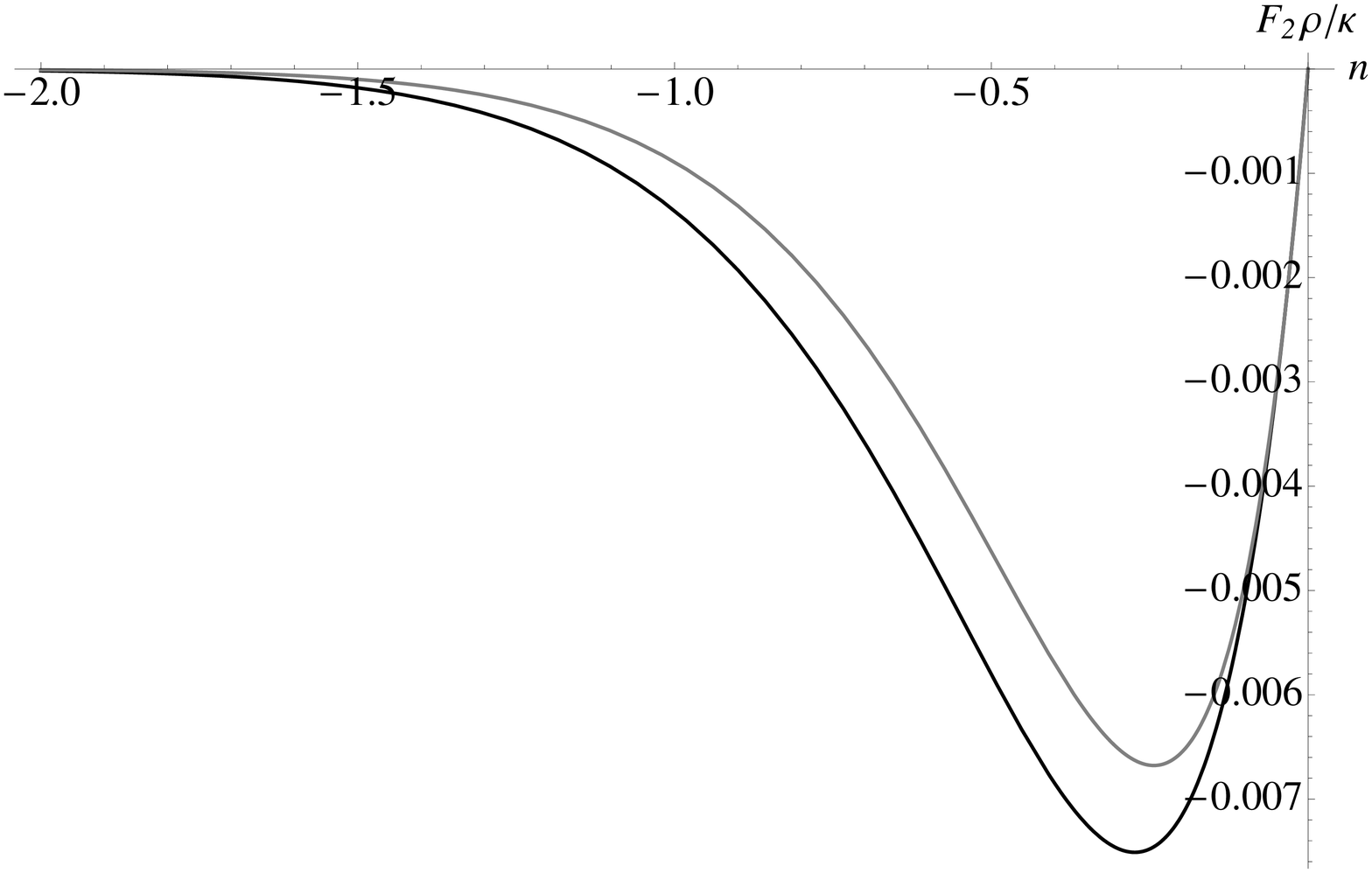}
\caption{Value of $F_2 \rho /\ka$ (\eq{cases}), assuming that $F_2\rho > \ka$ (black) or $F_2 \rho < \ka$ (gray).}
\label{whichgraf}
\end{figure}

\noindent The two curves above are depicted in Fig. \ref{whichgraf}, for negative exponent $n$: clearly, both are (in absolute value) always below unity, so that the condition $F_2 \rho > \ka$ is inconsistent, while the converse is valid for all $n$. Hence, one concludes that the $-$ regime $F_2 \rho < \ka $ is followed by the system once the non-minimal coupling becomes dominant.

\subsection{Discussion}

The previous section has allowed one to conclude that the effect of the non-minimal coupling leads to an evolution obeying condition $F_2 \rho < \ka$, so that the scale factor scales with a power-law $a(t) \propto t^\be(n)$, with $\be(n) = \be_-(n) \equiv 2(1-n)/3$. The assumption of accelerated expansion $q<0\rightarrow \be > 1$ (more stringent than the previously assumed $\be> 1$, where no acceleration is required) leads to the stronger constraint $n<-1/2$. One may resort to the expression for the deceleration parameter in \eq{scalarcurvature} and write instead

\beq \label{qn} n= 1- {3 \over 2(1+q)} \rightarrow q = -1 + {3 \over 2(1-n)} ~~. \eeq

\noindent Clearly, as $n \rightarrow -\infty$, one gets $q \rightarrow -1$: this is the expected value if one assumes that the effect of the non-minimal coupling mimics the $\La$CDM scenario, thus replicating a cosmological constant (although that implies an exponential evolution of the scale factor, not a power-law one). 

One may easily evaluate the equation of state (EOS) parameter $w$, as given by relation $p_c = w \rho_c$; manipulating the Friedmann \eq{friedmann} together with \eq{qeq} yields the well-known relation

\beq \label{omn} q = {1+3w \over 2} \rightarrow w = {2q-1 \over 3} = {n \over 1-n}~~. \eeq

\noindent Since the exponent $n$ is negative, one concludes that the EOS parameter of the mimicked dark energy obeys $ -1 < w < 0 $, thus fulfilling the weak, dominant and null energy conditions (for am extended discussion, see Ref. \cite{energy}).

As stated in Ref. \cite{mimic}, one may assume that the non-minimal coupling comprises several contributions, with a power-law referring to the dominant term (possibly of a Laurent series of a more evolved form) in a particular context. Hence, it is not required that the exponents $n$ of cosmological relevance play a part in astrophysical contexts, and vice-versa.

This said, the applicability of the model here considered to the puzzle of the flattening of the rotation curves of galaxies was studied in a previous work \cite{mimic}; in particular, it was found that the Navarro-Frenk-White and isothermal dark matter profiles can be derived from power-law non-minimal couplings with exponents $n_{NFW} = -1/3$ and $n_{IS} =-1$, respectively.

Notice that the constraint $n< -1/2$ rules out a cosmologically relevant non-minimal coupling $f_2(R)=(R/R_3)^{-1/3}$: this is in agreement with the results obtained in Ref. \cite{mimic}, where it was shown that the characteristic length scale $r_3 = 1/\sqrt{R_3}$ is much smaller than the relevant Hubble radius $r_H$.

Conversely, a simple inverse coupling $f_2(R) = R_1/R$ is allowed and yields an asymptotic deceleration parameter $q=-1/4$. From Ref. \cite{qz0}, one finds that this value lies still within the $2\si$ interval for the present value of $q(t)$; however, in Ref. \cite{mimic} it was found that, although such inverse coupling could play a cosmological role, it is not a dominant one (that is, the characteristic length scale $r_1 = 1/\sqrt{R_1} \lesssim r_H$).

Hence, one assumes that the $n=-1$ scenario does not correspond to the observed cosmological dynamics, and an even smaller value of the exponent $n$ is required (thus leading to a larger value of $|q|$): a fully consistent model would require that the cosmologically relevant coupling (characterized by an exponent $n_{C}$) does not disturb the dark matter mimicking scenario already obtained with the aforementioned exponents $n_{NFW}=-1/3$ and $n_{IS}=-1$. This shall be the object of a future study, as it is clearly outside the scope of the present work.

\section{Numerical results}

In this section, a numerical evaluation of the solution to \eq{qeq} is performed, varying both the time scale $t_2 = 1/\sqrt{R_2}$ and the exponent $n$. This equation is chosen, instead of the Friedmann \eq{friedmann} or Raychaudhury \eq{raychaudhuri} for three reasons: it best expresses the relation between a negative pressure and an accelerated expansion; it requires only the evaluation of $p_c$, not $\rho_c$; it allows for a direct comparison with available data.

In the context of this work, a natural candidate for an observable quantity is the evolution of the deceleration parameter with the redshift; since the proposed model offers a clear mechanism for the transition from the matter dominated phase (characterized by $q = 1/2$) to an accelerated expansion regime, one aims at comparing the solution to \eq{qeq} with available $q(z)$ evolution curves \cite{qz0} (see also Refs. \cite{alsoqz}), which employ fitting functions exhibiting an asymptotic future behaviour. It is useful to notice that these studies indicate that the deceleration parameter has not yet evolved completely to the permanent regime $q \rightarrow {\textrm const.}$; the best fit found for the present value of the deceleration parameter $q_0$ ranges from $-0.76$ to absolute values larger than unity.

With this considerations in mind, one may qualitatively predict the impact of varying the model parameters $t_2$ and $n$: increasing the former shifts the transition time $t_T \propto t_2$, {\it i.e.}, decreases the transition redshift $z_E$, defined by $q(z_E) = 0$ and found to lie in the range $0.2 < z_E < 0.4$. From \eq{qn}, decreasing the negative exponent $n$ lowers the asymptotic $q(n)$ value for the deceleration parameter.

At first glance, it appears that the problem of finding more suitable values for $n$ and $t_2$ is not too difficult, since one may ``guide'' these quantities to match the reported values for $q_0$ and $z_E$. However, the situation is somewhat more evolved: asides from obtaining the desired $q(z)$ profile (roughly specified by these quantities), one must also verify that the Hubble parameter $H$ and the matter density (or, equivalently, the scale factor $a(t) \propto \rho^{-1/3}$) acquire their present values. Furthermore, one should guarantee that the time assigned at the present does not deviate widely from $t_0$, {\it i.e.} $H(t_{now}) \simeq H_0$, $\rho(t_{now}) \simeq \rho_0$, with $t_{now} \simeq t_0$.

This said, the prospect of obtaining a completely coherent picture with only the postulated power-law non-minimal coupling appears unattainable. For this reason, one should clearly restate the purpose of this work: not to provide a thorough matching to the observational scenario, but to describe a possible mechanism through which a non-minimal gravitational coupling might account for the key features discussed: a transition from a matter dominated to a Universe with an asymptotic accelerated expansion. For this reason, two numerical studies were undertaken, as detailed below.

\begin{figure}[tbp]
\centering
\epsfxsize=\columnwidth
\epsffile{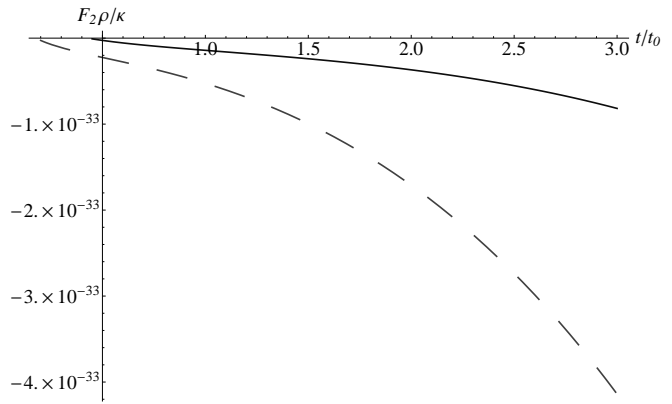}
\caption{Numerical result for $F_2 \rho /\ka$, for the first case ($n=-4$, $t_2 = t_0/4$, dashed) and second case ($n=-10$, $t_2=t_0/2$, full).}
\label{figregime}
\end{figure}

The validity of the $+$ regime was also ascertained, with $F_2 \rho / \ka$ found to lie below the $10^{-33}$ level. As can be seen in Fig. \ref{figregime}, this quantity is decreasing, instead of being constant, which does not contradict the finding of the previous section, since condition $F_2 \rho =  {\textrm const.} $ was derived in the context of a constant deceleration parameter, a regime which has not been attained yet.

For definitiveness, in Fig. \ref{figqz} one compares the solutions $q(z)$ to \eq{qeq} with the profile given by the fitting of the function

\beq q(z) = {1 \over 2} + {q_1 z + q_2 \over (1+z)^2}~~\eeq

\noindent to the Sloan Digital Sky Survey and the WMAP7 \cite{WMAP7} combined data, with best fit parameters $q_1 = 1.47$ and $q_2 = -1.46$ \cite{qz0} (Fig. 1 therein). The evolution of the EOS parameter $w$, derived from \eq{omn}, is presented in Fig. \ref{figwz}. Both figures show the solutions flowing to negative redshifts, thus showing the onset of the asymptotic regime.

\subsection{First case}

\begin{figure}[tbp]
\centering
\epsfxsize=\columnwidth
\epsffile{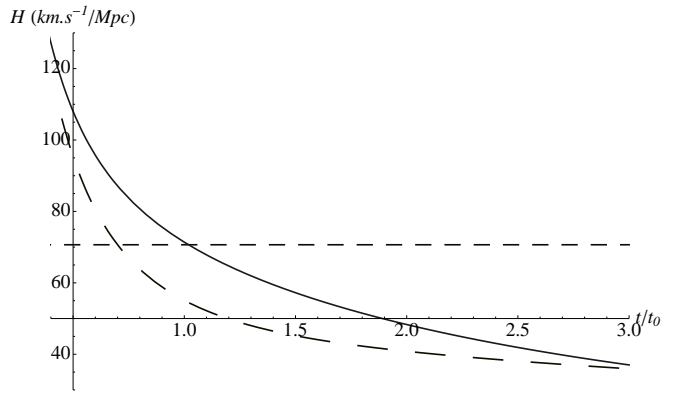}
\caption{Evolution of the Hubble parameter $H(t)$ for the first case ($n=-4$, $t_2 = t_0/4$, full) and the second case ($n=-10$, $t_2=t_0/2$, dashed).}
\label{figHt}
\end{figure}

\begin{figure}[tbp]
\centering
\epsfxsize=\columnwidth
\epsffile{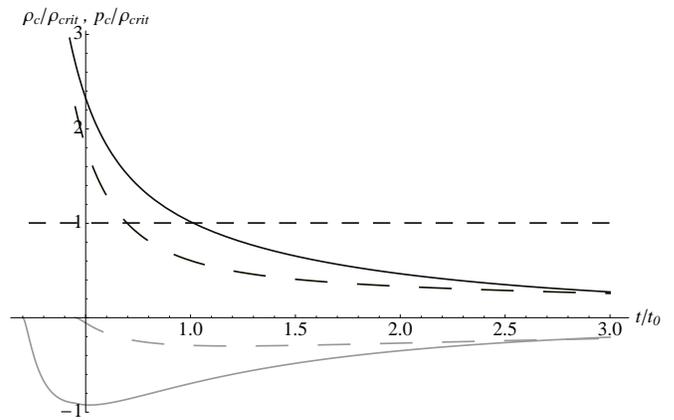}
\caption{Evolution of the curvature density $\rho_c$ (black) and pressure $p_c$ (gray) for the first case ($n=-4$, $t_2 = t_0/4$, full) and the second case ($n=-10$, $t_2=t_0/2$, dashed).}
\label{figrhot}
\end{figure}

This first numerical exploration aims at obtaining the evolution profile of the deceleration parameter $q(z)$, constrained by the requirement of matching the matter density  and the Hubble parameter, to their currently observed values at a time $t_{now} = t_0$ (as can be seen in Figs. \ref{figHt}, \ref{figrhot}).

By varying the exponent $n$ and the characteristic timescale $t_2$, it was found that this is obtained for $n= -4$ and $t_2 = t_0/4$. From Eqs. (\ref{qn}) and (\ref{omn}), one sees that the exponent $n = -4$ yields an asymptotic deceleration parameter $q(n=-4) = -0.7$ and an EOS parameter $w(n=-4) = -0.8$.

From Fig. \ref{figqz}, one sees that the agreement with $H_0$ and $\rho_0$ is attained at the expense of a higher value for $q_0 = -0.53 $ and the transition redshift $z_E = 1$. The solution $q(z)$ misses the indicated marks $q_0 \leq -0.76$ and $0.2 < z_E < 0.4$ and falls mostly within the $3 \si$ allowed region (except in the vicinity of $z = 0.3$); the region $z < 2$ falls within the $1 \si$ region.

\subsection{Second case}
The second attempt relaxes the above constraint, and aims instead to obtain a transition redshift within the range $0.2 < z_E < 0.4$, with most of the solution falling within the $1 \si$ allowed region. This is obtained for $n = -10$ and $t_2 = t_0/2$. From Eqs. (\ref{qn}) and (\ref{omn}), this value for the exponent yields an asymptotic deceleration parameter $q(n=-10) = -0.86$ and an EOS parameter $w(n=-10) = -0.91$.

\begin{figure}[tbp]
\centering
\epsfxsize=\columnwidth
\epsffile{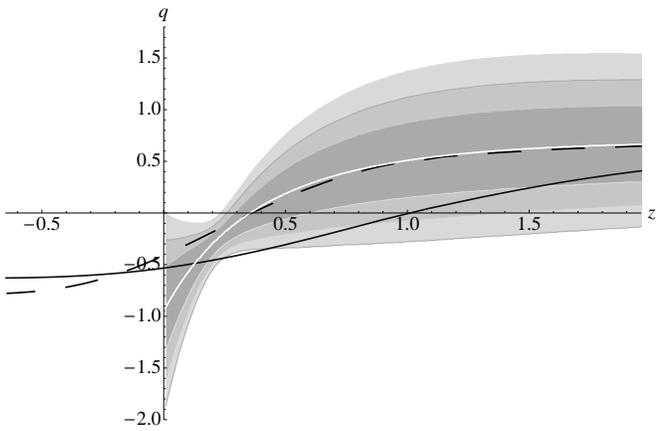}
\caption{Evolution of the deceleration parameter $q(z)$ for the first case ($n=-4$, $t_2 = t_0/4$, full) and the second case ($n=-10$, $t_2=t_0/2$, dashed); from Ref. \cite{qz0}, $1 \si$, $2\si$ and $3\si$ allowed regions are shaded, white line gives best fit.}
\label{figqz}
\end{figure}

\begin{figure}[tbp]
\centering
\epsfxsize=\columnwidth
\epsffile{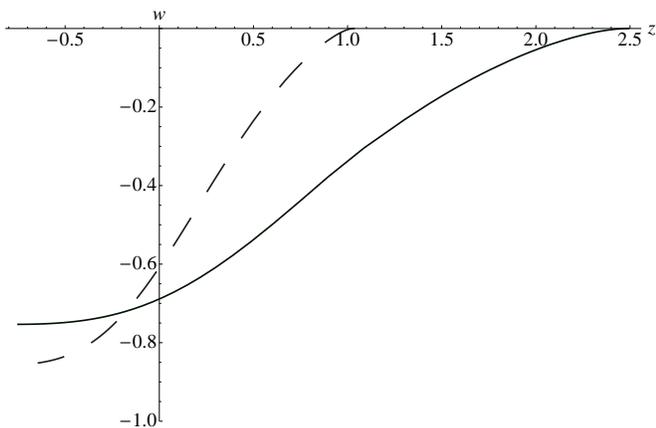}
\caption{Evolution of the EOS parameter $w(z)$ for the first case ($n=-4$, $t_2 = t_0/4$, full) and the second case ($n=-10$, $t_2=t_0/2$, dashed).}
\label{figwz}
\end{figure}

Following the previous discussion, Figs. \ref{figqz} shows the converse trade-off: the transition redshift $z_E = 0.36$ is obtained, and the solution $q(z)$ closely approaches the best fit curve before it, $z > z_E$; the current value for $q_0$ falls within the $2 \si$ region. However, both the Hubble parameter as well as the matter density deviate from their present values, $H(t_0) = 0.78H_0$ and $\rho(t_0) = 1.6 \rho_0$. 

Alternatively, one may express this mismatch by determining the times at which these quantities attain their present values: it is found that $H(0.7 t_0) = H_0 $ and $\rho (1.2 t_0) = \rho_0$. Clearly, this result does not amount to a simple shift of $t_{now}$ with respect to $t_0$, and expresses the aforementioned over determination of the observable quantities with respect to the available model parameters $n$ and $t_2$. 

\section{De Sitter Solution}

In the previous sections it was assumed that the Universe evolves towards an accelerated expansion phase due to a power-law non-minimal gravitational coupling $f_2(R)=(R/R_2)^n$, tailored from the assumed {\it Ansatz} for the scale factor $a(t) \propto t^{3\be}$. The latter does not allow for a De Sitter phase characterized by $q=-1$, which corresponds to an exponentially evolving scale factor $a(t) = a_0 \exp(H_0 t)$. From \eq{qn} it is found that this regime is approached in the limit $n \rightarrow -\infty$.

For completeness, one may instead investigate the particular form $f_2(R)$ that gives rise to a de Sitter spacetime. Since this yields a constant Hubble parameter $H = H_0$ and scalar curvature $R = 12 H_0^2$, the curvature density and pressure (Eqs. (\ref{rhoc}) and (\ref{pc}), respectively) take the form

\beqa \label{deSitter}
\rho_{c}&=&{\kappa\rho \over \kappa-F_{2}\rho}\left(f_{2}-30H_0^{2}F_{2}\right)~~,\\ 
p_{c}&=&{\kappa\rho \over \kappa-F_{2}\rho}6H_0^{2}F_{2}~~. \nonumber 
\eeqa

To ascertain the form for the non-gravitational coupling $f_2(R)$, one inserts the above relations into \eq{qeq}, obtaining

\beq q={1 \over 2} +{1 \over 4\ka} {p_c \over H_0^2} \rightarrow -1 = {F_2\rho \over \kappa-F_2\rho} ~~, \eeq

\noindent which has no exact solution, hinting that no dependence $f_2(R)$ will yield the result $q=-1$. Nevertheless, one may proceed and consider that the above equation hints that the non-minimal gravitational coupling $f_2(R)$ must follow the aforementioned ``strong'' $+$ regime, $F_2 \rho > \ka $.

It is trivial to argue that this regime cannot occur forever: the De Sitter phase is characterized by a constant scalar curvature {\it vis-a-vis} a constant $F_2(R)$; however, the matter density $\rho$ decreases monotonically, so that $F_2 \rho$ also drops. Hence, even if the condition $F_2 \rho > \ka$ is verified at the onset of the exponential expansion phase, it will eventually be untenable.

One must also check either the Friedmann \eq{friedmann} or the Raychaudhuri equation (\ref{raychaudhuri}), for consistency. From \eq{deSitter}, one has

\beq \rho_c + 3 p_c = {\ka \rho \over \ka - F_2 \rho} \left( f_2 - 12 H_0^2 F_2 \right) \simeq \ka \left( 12 H_0^2 - {f_2 \over F_2} \right)~~, \eeq

\noindent considering the previous condition $F_2 \rho > \ka$ for the $+$ regime.

Hence, the Raychaudhuri equation becomes

\beqa \label{difeq} 1 + f_2 = 24 F_2 H_0^2~~, \eeqa

\noindent which admits the solution 

\beq \label{dSsol1} 1+ f_2(R) = K \exp\left( { R \over R_2 } \right) ~~, \eeq

\noindent defining $R_2 \equiv 24 H_0$ and with $K$ an integration constant. This is clearly a unphysical result, since $K$ must be positive (so to match the Einstein-Hilbert action when $R \ll R_2$), one obtains an increasing function $f_2(R)$ of the scalar curvature. This becomes less and less relevant as $R$ decreases during the matter dominated phase, and the transition to the De Sitter phase never occurs --- on the contrary, the non-minimal coupling dominates at an early epoch.

For illustrative purposes, one entertains the following possibility: is it feasible to ``push the envelope'' in order to obtain a different form for $f_2(R)$, by reinterpreting \eq{difeq}? One may assume that the simple De Sitter phase is not actually enforced in nature, and the scalar curvature will not be constant: one could perhaps replace the factor $24H_0^2 \rightarrow 2R$, thus recasting \eq{difeq} as

\beq 1+f_2 = 2R F_2 ~~.\eeq 

\noindent However, this also yields an increasing solution $1 + f_2(R) = \pm \sqrt{R/R_2}$, with $R_2 $ a free parameter. Again, one must select the positive, increasing solution, with the same unphysical result.

Clearly, no amount of creativity can go against the simple interpretation of \eq{difeq}: since $1 + f_2$ must be positive, so should $F_2$. Hence, no form for the non-minimal gravitational coupling $f_2(R)$ yields a De Sitter phase.

In order to close this session, one may resort to a simple argument that corroborates this impossibility: the simplest model that originates a De Sitter phase relies on the presence of a cosmological constant. Hence, one may naively expect that the non-minimal gravitational coupling term should behave as a constant, $(1+f_2) \mathcal{L}_m \sim {\textrm const}$.

Recycling a previous point, one remarks that the constant scalar curvature yields a constant term $f_2(R)$, so that this mimicking of the cosmological constant would demand a non-evolving Lagrangian density $\mathcal{L}_m$! Neither the adopted choice $\mathcal{L}_m = -\rho$ or any of the classically equivalent ``on-shell'' forms ({\it i.e.} $\mathcal{L}_m = p$ \cite{Ld}) remain constant, since the relevant thermodynamic quantities decrease as the Universe expands.

\section{Conclusions}

In this work, one has applied a model exhibiting a non-minimal gravitational coupling with matter to the fundamental issue of the observed accelerated expansion of the Universe. In order to do so, one first assumes a constant deceleration parameter $-1< q < 0$: the related power-law form $a(t) \propto t^\be$ of the scale factor leads one to consider a power-law non-minimal coupling $f_2(R) =(R/R_2)^{-n}$, so that a negative exponent is required to drive the transition away from the early matter dominated phase.

In the analytical study, one first derived the equivalent form for the Friedmann and Raychaudhuri equations; since an exact solution does not exist, the two competing regimes $F_2 \rho > \ka $ or $F_2 \rho < \ka$ were discussed, and it was found that the later dominates throughout the evolution of the Universe. The identification $\be = 2(1-n)/3$ was thus obtained, as well as the dependence of the deceleration parameter $ q = - 1 + 3 / [ 2 ( 1 - n ) ] $ and the related EOS parameter $w = n/(1-n)$.

The obtained scenario was numerically tested through the variation of the parameters $n$ and $R_2$. The available evolution profiles for the deceleration parameter $q(z)$ and the EOS parameter $w(z)$ were compared with the obtained solutions, and it was found that a thorough fit of these results is not fully compatible with the current values for the Hubble parameter and the matter density. 

However, the driving force behind the present effort is not to yield a complete match with observations, but to thoroughly explore the proposed mechanism leading to an asymptotic accelerated expansion Universe. For this reason, the simplifying assumptions of a linear form for the pure curvature term $f_1(R) = 2\ka R$ was made, trading the perceived loss of flexibility that a combined analysis of non-trivial $f_1(R)$ and $f_2(R)$ might yield with the advantage of deriving analytical results.

With this in mind, the present work should be regarded as a first step towards a more complete description based on the encompassing concept of modifying the Einstein-Hilbert action, and in a similar way to what was performed regarding the flattening of the galaxy rotation curves, a future improvement would include the effect of a combination of power-laws for $f_2(R)$. More ambitiously, one could attempt to reverse-engineer the exercise and read the form for $f_1(R)$ and $f_2(R)$ from the observed evolution profiles, using the latter as inputs, instead of targets.

Despite the above justification, one could consider that the results that were used for comparison (based on Ref. \cite{qz0}) employ a fitting function that might be proven inadequate. Indeed, the main qualitatively difference between these scenarios and the solutions obtained in this work is that the latter present a much smoother transition from the matter dominated to the accelerated expansion phase: speculatively, one can state that perhaps the relatively small number of well selected supernovae observations close to the recognized transition redshift $ z_E \sim 0.3$ might allow for a shallower $q(z)$ transition.

A final remark is in order: earlier in the text, it was suggested that the versatility of the proposed model naturally accounts for 
the effects of the presence of the two ``dark'' components of the Universe: dark energy and dark matter. Indeed, the possibility that different terms present in the non-minimal coupling $f_2(R)$ manifest themselves at distinct scales, astrophysical or cosmological in nature, does not require that the same exponent $n$ and characteristic scale $R_n$ is shared between widely differing phenomena.

In the present case, it was shown that the $n=-1$ or $n=-3$ exponents, which have been shown to account for the observed flattening of the galaxy rotation curves, do not have cosmological relevance --- allowing for other values for $n$ to assume such role. Notwithstanding, it is interesting to notice that the same EOS form arises both for dark matter as well as for dark energy, given by the parameter $w = n/(1-n)$.

This unification is also illustrated by the formulation of the proposed model as a multi-scalar-tensor theory \cite{scalar}: in this context, an interesting target for future research lies in the possibility of bridging the mechanism here described with others encountered in the literature, namely quintessence models \cite{quintessence}, chameleon fields \cite{chameleon} or, the aforementioned generalized Chaplygin gas unified model \cite{Chaplygin}.

\end{document}